# SUPERCONDUCTING PROPERTIES OF HIGHLY DENSE MgB$_2$ BULK MATERIALS

G.GIUNCHI, S. CERESARA

*EDISON S.p.A. – Divisione Ricerca e Sviluppo – Via U. Bassi 2  20159 Milano (Italy)*
*E-mail: giovanni.giunchi@edison.it*

L.MARTINI, V. OTTOBONI

*CESI SpA, Via Rubattino 54, 20134 Milano, Italy*

S.CHIARELLI, M.SPADONI

*ENEA - Centro Ricerche Frascati – via E.Fermi 45  00040 Frascati (RM)*

Highly dense MgB$_2$ policrystalline bulk materials, obtained by reactive liquid infiltration, have been characterized in their superconducting transport and magnetic properties in magnetic field and in a temperature range of interest for the MgB$_2$ compound. The products have a granular morphology of composite nature and their overall density affects their properties. The irreversibility line, B$_{c1}$(T) values, magnetically measured current densities and the trapped magnetic field are evaluated by susceptibility curves and hysteresis curves, both obtained by SQUID magnetometry for samples of different density, up to 35K and up to 5T applied magnetic field. The experimental results have been compared with the best values reported in the literature for other bulk sintering techniques.

## 1  Introduction

The sintering of the bulk MgB$_2$ material is generally performed with high pressure apparatus to avoid the decomposition, with Mg release, that occurs at high temperature. Two main approaches up to now have been followed for the preparation of the bulk manufacts: the synthesis from the elemental powders (defined "in-situ") and the sintering of already formed MgB$_2$ powders ( defined "ex-situ"). A variant process of the in-situ like technology has been recently introduced [1]. According to this new technology the liquid Mg infiltrates a porous perform of B powders and from their reaction very dense MgB$_2$ is produced. The granularity of the resulting material is not detrimental to the flowing of the superconductive currents [2] and consequently interesting applicative properties can be foreseen for the bulk manufacts so obtained.

In this paper we presents the first superconductive characterization of the material obtained by such a process. In particular we show the effect of the resulting density of the manufacts on its superconductive properties, such as the critical current density and the irreversibility field. A comparison with bulk MgB$_2$ obtained with other technologies is also given. The range of the examined temperatures, up 35 K, and of the applied magnetic fields, up to 5 T,  give insight on the potentiality of the material for the power applications. The main potential advantages are the use of a refrigeration system based on cryocoolers, less energy consuming with respect to the use of  liquid helium, and the realization of medium high magnetic fields , larger than the fields obtained by conventional electromagnets.

## 2  Samples preparation

The general description of the MgB$_2$ preparation procedure, based on the reactive liquid Mg infiltration of B powders, is described elsewhere [3]. Here we consider a similar preparation of a new sample , with the same quality of the starting materials but slightly different operative conditions with respect to the sample (A) of the said paper, that we take here as reference sample, again referred in the following as sample (A). For the new sample, referred in the following as sample (B), use has been made of an external cylindrical stainless steel container of 1.75 mm wall thickness (4 mm for sample (A)), with a slightly higher Mg/B ratio , i.e. a molar ratio of 0.649  ( 0.633 for sample (A)). The sealing conditions of the steel container in inert atmosphere and the thermal treatments were equal for the two samples, namely 950°C x 3 hours. Figure 1 shows the morphologies of the two samples , both characterized by apparent large grains embedded in intergrain zones with finer grains. Sample B shows a larger proportion of the intergrain zone, with respect to sample A.

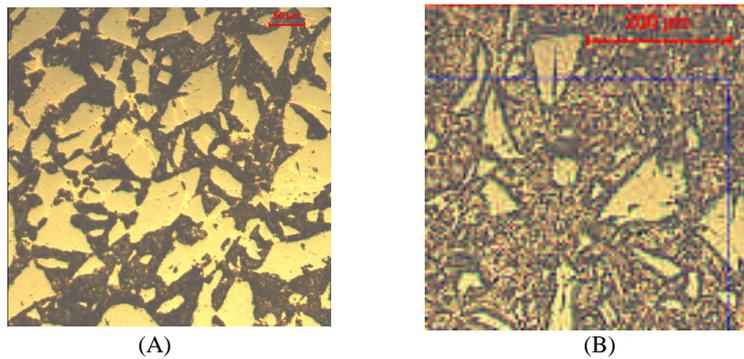

(A)                                        (B)

**Figure 1** – Optical micrographs of the two examined $MgB_2$ bulk samples: (A) previous preparation ; (B) present preparation

The large grain material is not single-crystalline but has a polycrystalline nature similar to that present in the intergrain zone, as evident by the SEM analysis shown in Figure 2. The backscattered electrons microanalysis tell us that the major difference between intergrain and large grain zone is the amount of residual Mg,  with a larger Mg amount in the former.

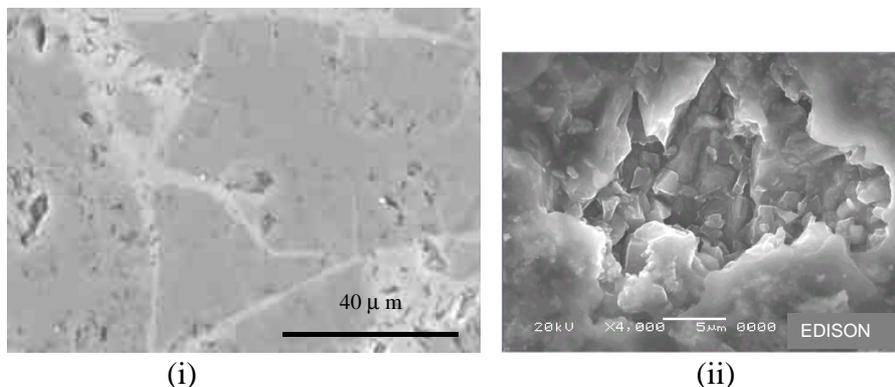

(i)                                        (ii)

**Figure 2** – SEM analysis of the large grain morphology of the sample (A): (i) backscattered electrons image of the large grains ; (ii) an image at higher magnification of an hole inside the large grains reveals a finer grain morphology

The density of the two samples is slightly different , going from 2.4 $g/cm^3$ of sample A  to 2.3 $g/cm^3$ of sample B. In any case, it is much higher than the values reported for low pressure sintering techniques [5]. A precise comparison with the theoretical density of the $MgB_2$ (2.63 $g/cm^3$) cannot be performed due to the surplus of Mg remaining in the samples .

## 3  Critical current transport measurement

The measurement of the critical current density has been performed with the standard 4 point technique at 4.2K for sample A. From the original $MgB_2$ pellet, a bar of dimensions (1 x 6 x 30 $mm^3$) has been cut. The joining of the bar to the current leads and to the voltage intake has been performed with silver paste. The E(I) characteristics have been measured only at high magnetic fields , due to the contact resistance and are shown in Figure 3. The critical current density has been determined according to the 1μV/cm criterion. The n parameter of the $E\sim I^n$ interpolating equation is varying from 15.2 at B=9 T to 2.6 at B=12T

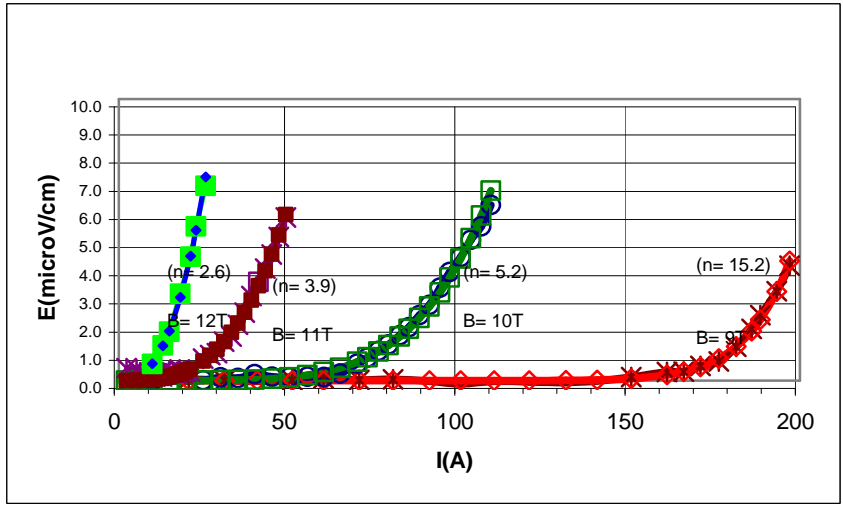

**Figure 3** – E(I) characteristics of the sample A , at 4.2 K, in different magnetic fields.

As commented elsewhere [3] the values of the critical current density of the bar A , displayed in Figure 4, are the best up to now published for bulk $MgB_2$ material.

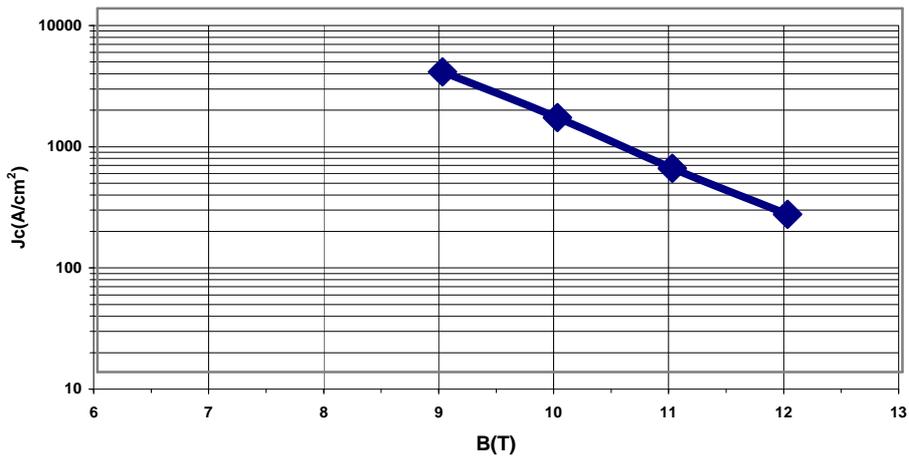

**Figure 4** – Critical current density, as function of the magnetic field B, measured by transport method a 4.2 K

## 4  Magnetic characterization

In order to compare the superconductive properties of the two samples, at higher temperatures, we have performed a SQUID magnetometer analysis in magnetic field up to 5 T and at temperatures up to 35 K(Oxford Instruments). Both samples have been cut to a bar of final dimensions of 0.95x1x 6.8 mm$^3$ and are inserted in the magnetometer with the long axis along the magnetic field.

### *4.1  Hysteresis magnetization cycle*

The magnetization loops of the two samples are shown in Figure 5. The sample A has in general better values at the various temperatures where comparison can be done. But at lower temperature (4.6 K) the sample A shows a more pronounced instability, represented by the glitches in the lower applied fields part of the loop. The field associated to the departure from the linearity in the initial M(H) curve is defined as the lower critical field, $B_{c1}$, which separate the complete diamagnetic Meissner-like behaviour of the type I superconductors from the mixed state behaviour typical of the type II superconductors. In our case the mean value of Bc1 is similar for the two samples, with a value of about 40 mT at 5 K and about 35 mT at 20 K. The evaluation of the critical current density can be

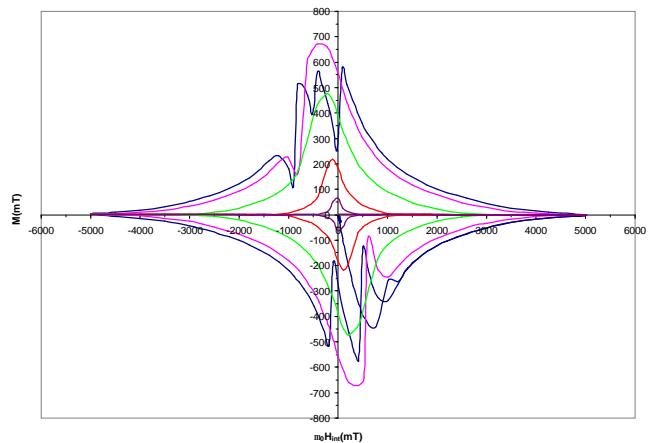

**Figure 5** - Magnetization loops for sample A and sample B at various temperatures.